\documentclass{article}

\usepackage[svgnames, x11names]{xcolor}

\newcommand{\ysnoted}[1]{}

\usepackage[normalem]{ulem} 

\usepackage{listings}

\lstdefinelanguage{XML}
{
basicstyle=\ttfamily\footnotesize,
  morestring=[b]",
  moredelim=[s][\bfseries\color{Maroon}]{<}{\ },
  moredelim=[s][\bfseries\color{Maroon}]{</}{>},
  moredelim=[l][\bfseries\color{Maroon}]{/>},
  moredelim=[l][\bfseries\color{Maroon}]{>},
  morecomment=[s]{<?}{?>},
  morecomment=[s]{<!--}{-->},
  commentstyle=\color{gray},
  stringstyle=\color{blue},
  identifierstyle=\color{red}
}
\usepackage[pdftex]{graphicx}
\graphicspath{{./figures/}}
\DeclareGraphicsExtensions{.pdf}

\usepackage[cmex10]{amsmath}
\usepackage{amssymb}
\usepackage{mathtools}

\usepackage[]{subfig}

\usepackage{algorithmicx}
\usepackage{algpseudocode}
\usepackage[ruled]{algorithm}
\definecolor{light-gray}{gray}{0.75}
\algrenewcommand{\algorithmiccomment}[1]{\hskip3em{{\footnotesize \textcolor{light-gray}{$\blacktriangleright$}}} #1}

\usepackage{multirow}

\usepackage{hyperref}

\usepackage{xspace}
\usepackage[nocompress]{cite}

\usepackage{enumitem}

\usepackage[english]{babel}
\usepackage{blindtext}

\begin{document}
%
\title{Demystifying Fog Computing: Characterizing Architectures, Applications and Abstractions}
%
%
%
%

\author{Prateeksha Varshney and Yogesh Simmhan \\
\normalsize{\emph{Department of Computational and Data Sciences}}\\
\normalsize{\emph{Indian Institute of Science (IISc), Bangalore 560012, India}}\\
\normalsize{\emph{Email: prateeksha@grads.cds.iisc.ac.in, simmhan@cds.iisc.ac.in}}}

\date{}
\maketitle

\begin{abstract}
	Internet of Things (IoT) has accelerated the deployment of millions of sensors at the edge of the network, through Smart City infrastructure and lifestyle devices. Cloud computing platforms are often tasked with handling these large volumes and fast streams of data from the edge. Recently, Fog computing has emerged as a concept for low-latency and resource-rich processing of these observation streams, to complement Edge and Cloud computing. In this paper, we review various dimensions of system architecture, application characteristics and platform abstractions that are manifest in this Edge, Fog and Cloud eco-system. We highlight novel capabilities of the Edge and Fog layers, such as physical and application mobility, privacy sensitivity, and a nascent runtime environment. IoT application case studies based on first-hand experiences across diverse domains drive this categorization. We also highlight the gap between the potential and the reality of Fog computing, and identify challenges that need to be overcome for the solution to be sustainable. Together, our article can help platform and application developers bridge the gap that remains in making Fog computing viable.	
\end{abstract}
\footnote{Pre-print: To appear in ICFEC-2017 proceedings}


\section{Introduction}

The prime drivers of Big Data over the past decade have been the WWW, social media, eCommerce, and enterprise information systems, with data and services being consolidated in public and private data centers. However, \emph{sensing at the edge} of the network is accelerating many-fold. This is enabled through infrastructure like Smart Cities that are being deployed globally, lifestyle devices like wearables and smart appliances, and observations collected \emph{ad hoc} through crowd-sourcing over smart phones~\cite{satya:pervasive:2015,aman:comm:2013}. A large class of this data streaming from the edge can be attributed to the \emph{Internet of Things (IoT)}, where the convergence of commodity sensors, accessible communications, and the need for intelligent infrastructure management is driving this exponential growth.

The traditional model of acquiring and processing this data from the edge to offer useful services has been through \emph{Cloud computing}. These large Cloud data centers host computing and storage infrastructure with high bandwidth connectivity, and allow IoT applications and services to be hosted remotely~\cite{yannuzzi:camad:2014}. Their on-demand access to seemingly infinite resources accessed through simple web service APIs, combined with the pay-as-you-go billing model that capitalizes on economies of scale have made Cloud computing popular for supporting millions of clients at the edge.

At the same time, two key challenges emerge with this hub-and-spoke model. Public Cloud data centers are globally distributed, but typically limited to one or a few per country (depending on the size). This means that the network distance from the edge to the Cloud results in round trip time (RTT) are in the order of $100's$ of milliseconds, just in terms of latency~\cite{satya:pervasive:2009}. Many IoT applications are \emph{latency sensitive}, be they demand prediction for Smart Power Grids~\cite{aman:smartgridcomm:2015,interrante:hpcc:2012} or voice responses in Apple Siri. Consequently, hosting the analytics and decision making in the Cloud can compromise their performance~\cite{dastjerdi:computer:2016,Bonomi2014}. 

Similarly, video is emerging as a major source of data from the edge, from urban surveillance and ATM cameras that are part of the infrastructure, to personal body cameras and crowd-sourced recordings from phones~\cite{satya:pervasive:2015,yi:mobidata:2015}. Here, the \emph{bandwidth} required to push high fidelity video streams to the Cloud is prohibitive, and this further increases the RTT if the video analytics on the Cloud has to control edge devices, like zooming a PTZ Camera or detecting faces in real-time. 

The has led to a design paradigm of processing at the edge to supplement, or replace, Cloud computing~\cite{bittencourt:pgcic:2015,ghosh:arxiv:2016}. Also called  \emph{Edge Computing} or \emph{Mobile Cloud}~\cite{yi:mobidata:2015,yannuzzi:camad:2014}, these are typically used in vertically integrated applications where a part of the processing and analytics happens on the edge device while the Cloud is used for coordination and data archival. For e.g., a FitBit fitness watch may pre-process and visualize data on the smart phone before archiving to the Cloud.

However, Edge computing suffers from several deficiencies. The edge platforms tend to be constrained devices, with battery capacity or memory often being the limiting factor rather than even compute capability. This can cause resource contention if multiple IoT applications need to be deployed concurrently~\cite{dastjerdi:computer:2016}. Also, there are no well-defined or robust runtime or management platforms for composing generic Edge computing applications, comparable to the virtualization or service-based architectures exposed in Cloud computing. As a result, edge computing is limited to bespoke solutions.
%

\emph{Fog computing}~\cite{Bonomi2014}, also known as \emph{Cloudlets}~\cite{satya:pervasive:2009}, were introduced by Satyanarayanan, et al., and popularized by Cisco as a complementary resource-rich layer that sits between the edge device and the Cloud. \cite{satya:pervasive:2009} suggest that this layer be one network hop away from the edge to offer both low and predictable latencies to support gaming and video conference services. The Cloudlets are meant to serve as small-scale data centers that are placed closer to the edge~\cite{bittencourt:pgcic:2015}, where often both the generators of the observations and the consumers of the analytics reside. At the same time, the Fog layer can offer easier manageability of resources as services, and possibly a feasible business model similar to the popular Clouds.

In this regard, Fog computing has similarities with \emph{Content Distribution Network (CDN)}, that sit close to the edge for low latency delivery of content, typically static or slow changing and populated from the Cloud, but are otherwise passive hosts. \cite{satya:pervasive:2015} considers Cloudlets as a ``CDN in reverse'', with the edge populating the Cloudlet, and the Cloudlet serving data to the Cloud. At the same time, the Fog can also actively host services and analytics -- closer to the edge than, even, CDN.

From a societal perspective, Fog computing also becomes relevant when there are network outages that restrict or remove the ability to reach the Cloud data center. For e.g., the \emph{Cyclone Vardah} that hit Chennai, India in Dec. 2016 damaged a trans-oceanic optical cable connecting India with Singapore and Europe, where many commercial data centers are located~\footnote{A huge storm has messed up India's Internet by Rishi Iyengar, CNN, December 14, 2016 \url{http://money.cnn.com/2016/12/14/technology/india-cyclone-vardah-chennai-internet/}}. In future, as critical services such as water, power and transport are controlled by smart algorithms, hosting them in a Fog layer with peering mechanism allows a graceful degradation of functionality even in the absence of access to the Cloud, rather than cause a debilitating loss of city services.

However, it is important neither to over-state the potential of Fog computing nor to dismiss it as hype. Despite the existence of the concept for several years now, commercial deployments of Fog Computing are yet to take off. Part of the reason is that there are still an inadequate number of widely deployed or critical applications that find the Fog to be essential. There is a also lack of clarity on the application model, runtime and management environments for a Fog platform. And lastly, a sustainable business model and service providers are still evolving. So it is useful to take a \emph{critical look} at Fog computing as a \emph{prospective} pervasive platform. 

In this paper, we attempt to characterize some of the key features of the system design and computing architecture of Fog computing; the application features, particularly from IoT use-cases, that motivate the need for Fog computing; and lastly the programming models and abstractions that can be leveraged to bridge the gap between the applications and the systems for designing intuitive runtime platforms.

Several papers offer desiderata for Fog computing and its concepts. Many tend to be superficial or narrowly defined. Cisco subsumes the Fog into the Edge (or vice versa), there by discounting widely-available devices like smart phones and Raspberry Pis deployed as part of Smart City infrastructure~\cite{Bonomi2014}. 
\cite{vaquero2014finding} attempts to define Fog computing, but views it from the limited prism of network management and connectivity. Several papers on Cloudlets offer exemplar applications that are actually deployed, but fail to offer an overarching technical view of this space~\cite{satya:pervasive:2009,satya:pervasive:2013,satya:pervasive:2015}. Others propose a programming model for composing applications that run across mobile, Fog and Cloud layers as a Platform as a Service (PaaS), but limit it to a rigid hierarchy~\cite{hong:mcc:2013}. \cite{luan:arxiv:2016} attempts a similar effort as our paper, but favor mobile devices rather than edge devices at large. The latter, many of which can be part of an IoT infrastructure, are key. \cite{dastjerdi:computer:2016} discuss the role of Fog computing on IoT applications, and highlight open issues. These are further discussed in the related work section (\S~\ref{sec:related}).

Here, however, we go a step further and provide intrinsic dimensions and capabilities of Fog computing that should be of interest to platform developers and application designers, while recognizing that this is still an emerging space. These features distinguish the Fog from traditional mobile or Cloud computing but \emph{form a continuum}. These characteristics are as yet inadequately studied in the context of a holistic Edge, Fog and Cloud computing eco-system. We also motivate the problem space with emerging, rather than futuristic, applications.



We make the following specific contributions in this paper:
\begin{enumerate}
	\item We characterize the \emph{expected capabilities of a Fog computing system}, and lay them in the context of existing Edge/Mobile and Cloud computing architectures. We highlight dimensions that distinguish these alternative, but inter-linked, resource layers (\S~\ref{sec:arch}).
	
	\item We review the \emph{application requirements} that motivate the need for a Fog layer, and identify the gaps posed by a Cloud-only or Edge+Cloud model. We also offer case studies of applications and their features (\S~\ref{sec:apps}).
	
	\item We highlight the possible \emph{runtime and middleware capabilities, and application models} required in a Fog layer, 
	while recognizing that the Fog platform will need to adapt to the unique needs of applications and 
	the actual manifestation of the Fog system (\S~\ref{sec:platform}). 
	
	\item Lastly, we \emph{discuss} the gap between potential and reality of Fog computing, and identify challenges that need to be overcome and solutions that are possible in this evolutionary phase (\S~\ref{sec:discuss}).
	
\end{enumerate}


\section{Fog Computing System Architecture}
\label{sec:arch}


\subsection{Definitions}
There exist various overlapping definitions of what Fog computing is, in literature. We first summarize some of these views, and then take up a more detailed characterization.
\begin{itemize}
	\item Satyanarayanan, et al, \cite{satyanarayanan:2009} state that Cloudlets, synonymous with Fog computing~\cite{satya:comm:2015}, are a resource-rich (cluster of) computers that are located one hop away from the mobile devices. They have Gigabit interconnect, and high bandwidth, through Wireless to the mobile device and through WAN to the Internet. The Cloudlets offer a ``data center in a box'' close to the mobile device in order to reduce the network latency and bandwidth-induced latency to support interactive applications.
	
	\item Cisco~\cite{Bonomi2014} views Fog Computing as offering resources like compute, storage and networking similar to Clouds, with support for virtualization and multi-tenancy. These are geo-distributed to offer low and predictable latencies to client applications that are mobile or part of the infrastructure. They do conceive of the Fog as having modest resource capabilities, ranging from an edge network router to a high-end server.
	
	\item \cite{vaquero2014finding} define Fog computing as a large collection of heterogeneous and decentralized devices, communicating among themselves to store data and process tasks, that can be leased to users to support basic network functions or sandboxed applications. They take a communication-centric view of Fog computing, with 4G/5G connectivity enabling the Fog layer and Software Defined Network (SDN) management pushed to the Fog.

\end{itemize}

\subsection{Comparing Edge, Fog and Cloud Computing}
From the above definitions, among many others, Fog computing is seen as a resource layer that fits between the edge devices and the Cloud data centers, with features that may resemble either. Hence, it is worth comparing and contrasting the characteristics of the Fog computing system architecture relative to the capabilities of Edge and Cloud computing.

\begin{figure}[t]
	\centering
	\includegraphics[width=\columnwidth]{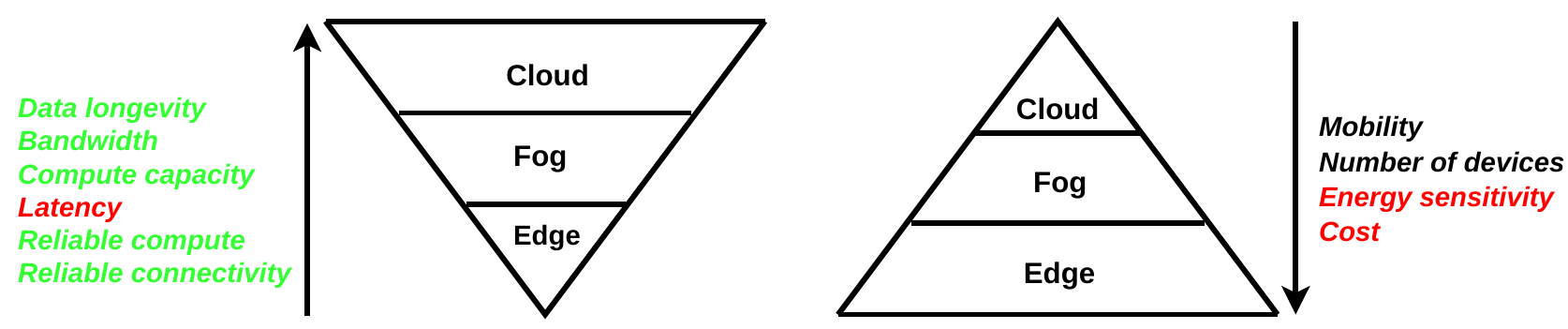}
	\caption{Resource characteristics of Cloud, Fog and Edge computing systems}
	\label{fig:pyramid:1}
	\vspace{-0.15in}
\end{figure}
\subsubsection{Resource Characteristics} There are several resource and performance characteristics that distinguish these three layers, as captured in a pyramid structure in Fig.~\ref{fig:pyramid:1}. Chief among them, serving as the fundamental motivation, are the \emph{network characteristics}. The Fog is close to the edge in the network topology. Hence, it has a lower latency for access in \emph{both} directions, i.e., serving content from Fog to Edge, and pushing data from Edge to Fog~\cite{satyanarayanan:2009}. Whether it is at a 1-hop or multi-hop depends on the deployment. Further, the bandwidth between the Edge and Fog is also higher (e.g., using WiFi) than from the Edge to the Cloud (e.g., using cell networks). The Fog is also expected to have a high-bandwidth and reliable link to the Internet~\cite{stojmenovic:2014}. The connectivity between the Edge and the Fog may be less robust than between the Fog and the Cloud due to the use of wireless links for the last mile~\cite{luan:arxiv:2016}.

In these regards, the Fog layer is analogous to a CDN but typically closer to the Edge~\cite{vaquero2014finding}. The \emph{storage capacity and data longevity} of Fog layer is much higher than the Edge devices, though more limited than Clouds. This storage can be used to cache large datasets that are useful to edge devices, such as apps or Virtual Machine (VM) images that have to be installed on devices for Smart City infrastructure, or snapshots of historic data~\cite{bittencourt:pgcic:2015}. But more interestingly, the Fog serves as a ``reverse CDN'' to allow edge devices to push data to the Cloud~\cite{bonomi2012fog}. This allows scenarios where data is staged in the Fog and periodically pushed to the Cloud for archival, potentially after some pre-processing such as deduplication~\cite{satya:pervasive:2015,dasterdi:corr:2016}. Further, edge devices can also access data pushed up to the Fog by other edge devices as well~\cite{luan:arxiv:2016}.

At the same time, the Fog layer also offers \emph{compute resources} that have a higher capacity and reliability than the Edge but to a smaller scale than Cloud data centers~\cite{stojmenovic:fedcsis:2014,Bonomi2014}. 
This resource capability can be used to host applications and services that range from Video Analytics as a service for processing video frames close to the Edge or directory services where devices register and access location-sensitive configuration for mobile phones~\cite{Loke2015TheIO}.

The sheer number of Fog appliances will dwarf the number of Cloud data centers (even if not their cumulative compute power), just as the number of edge devices number in the billions. This makes \emph{manageability} of the Fog more challenging, but easier than the Edge. The economies of scales will also come into play if Fog hardware is standardized, similar to commodity smart phones or shipping containers with Cloud hardware. This can make Fog computing affordable and (if the edge device is not captive) cheaper than Edge computing, though Clouds will retain their \emph{pricing} advantage. 

Lastly, the \emph{energy profile} can influence the capability and availability of some resources. Edge devices are often concerned with battery life, and the choice of using specific Edge features may depend on current battery level~\cite{mishra:iot:2015}. Cloud data centers reduce their energy footprint, but to limit operational costs~\cite{bittencourt:pgcic:2015}. The Fog layer is expected to run off grid power and, like the Cloud, be conscious of energy use to ensure fair pricing~\cite{vaquero2014finding,Loke2015TheIO,dasterdi:corr:2016}. But there may be cases of remote deployments where the Fog may run off renewables like solar where energy conservation may be a primary goal.



\begin{figure}[t]
	\centering
	\includegraphics[width=0.7\columnwidth]{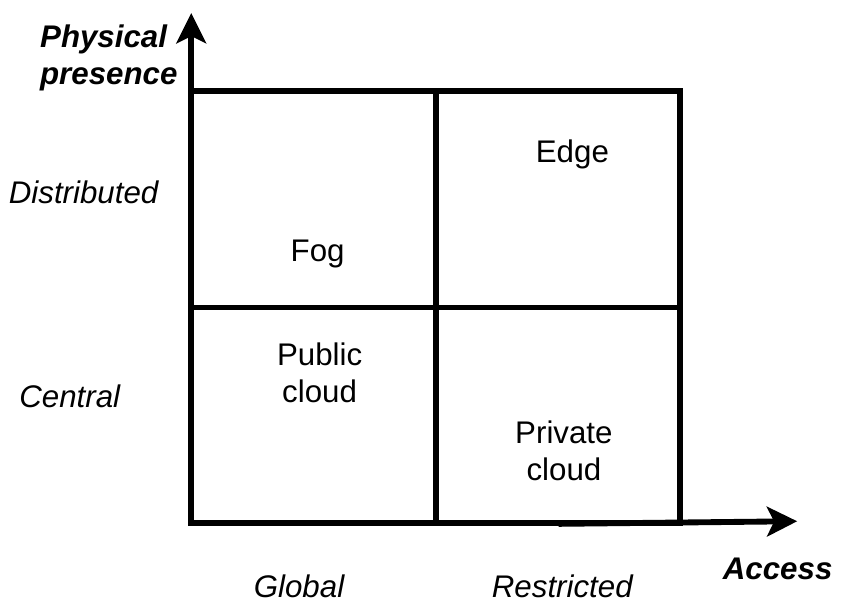}
	\caption{Physical presence vs access of Cloud, Fog and Edge resources}
	\label{fig:physical}
	\vspace{-0.15in}
\end{figure}
\subsubsection{Physical Presence and Access} Besides network distance, it is also worth considering the physical distance between the three computing paradigms, and their accessibility by clients. In Fig.~\ref{fig:physical}, four quadrants are formed from considering whether the resources within a layer are \emph{physically centralized or distributed (Y Axis)}, and whether their \emph{access is global or restricted (X Axis)}.


Resources in a Cloud data center are centrally located, but depending on whether the Cloud is public or private, are available to any client in a pay-as-you-go model, or only to users who are part of the private corporation~\cite{luan:arxiv:2016}. That said, public Cloud providers host multiple data centers that are geographically distributed, sometimes several in a country or continent, while the number of large private data centers for an enterprise tends to be more limited.

We can compare Edge and Fog resource distribution and access along a similar scale. Edge resources such as smart phones and set top boxes are distributed far and wide, but the access to them are restricted to individual users or managed applications~\cite{Bonomi2014,yi:mobidata:2015}. Fog resources are also geographically distributed to be close to the edge, but not as dispersed though much more than Cloud data centers. Additional specializations, on whether there is a Fog for each city block, one for the whole city or other variants, depend on the business models and applications that will evolve. One also expects the Fog to offer as a shared, pay-as-you-go IaaS or PaaS model~\cite{bittencourt:pgcic:2015,yi:mobidata:2015}.

The distributed nature of Edge and Fog resources increases their \emph{attack and failure surface}. There is a high chance of some device or Fog server failing or a network link dropping, and resiliency has to be built into the platform and application if necessary~\cite{madsen2013reliability}. Clouds being more centralized are single points of failures, but 
Cloud fabrics and Big Data platforms internalize faults within the data center. But, as mentioned, natural disasters can cause these central hubs to get islanded, and bugs and security breaches can cause massive data loss.

The access restrictions on private Clouds and Edge devices translates to a \emph{zone of trust} for applications and services hosted on them, which enables sensitive data and services to be hosted on them. Fog and public Clouds, however, are designed as shared resources with \emph{multi-tenancy}, which require higher measures of security and sand-boxing between different applications or users, using containers or hypervisors~\cite{vaquero2014finding,dastjerdi:computer:2016}.

That said, there may be Fog architectures where the resources are deployed for specific applications or organization (e.g., a Smart City municipality) and not available for rent, similar to a private Cloud~\cite{luan:arxiv:2016}. Further, the Fog layer may sit at the boundary between the public and private network. For e.g., as part of the IISc Smart Campus project~\footnote{IISc Smart Campus Project, \url{http://smartx.cds.iisc.ac.in}}, there are hundreds of edge devices and sensors connected to the campus LAN that monitor the water, power and road network, and Fog servers on the DMZ are connected both to the private LAN and to the public WAN. This makes them well-suited to run proxy services that translate from one zone of trust to another, one service layer to another (e.g., CoAP to HTTP), or one network protocol to another (e.g., IPv6 to IPv4).

\begin{figure}[t]
	\centering
	\includegraphics[width=0.8\columnwidth]{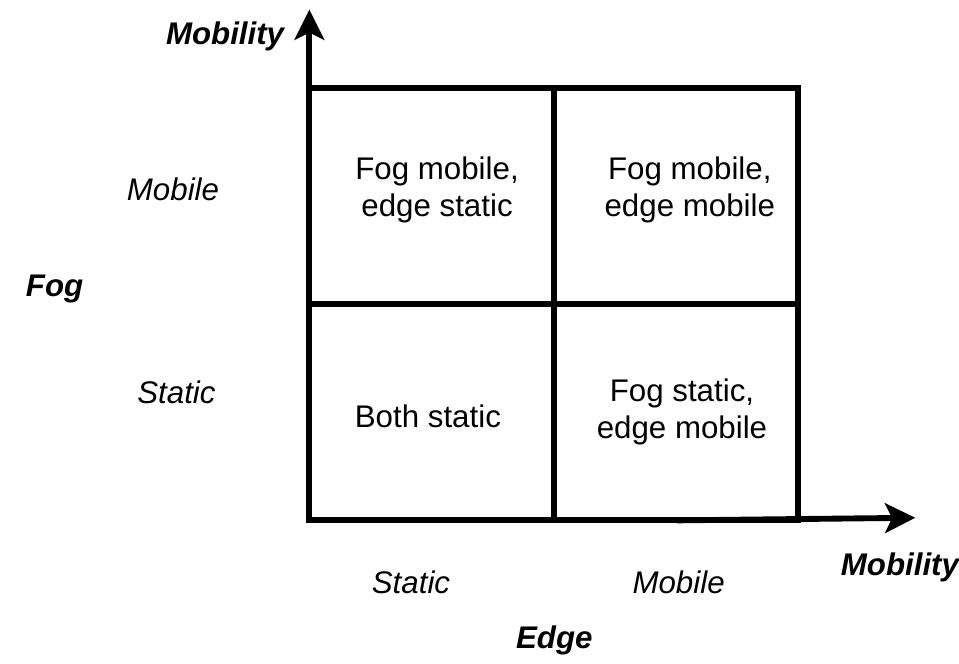}
	\caption{Physical mobility of Fog and Edge resources}
	\label{fig:mobility}
	\vspace{-0.15in}
\end{figure}
\subsubsection{Mobility}
It helps to understand the impact of mobility on these three resource layers as this impacts the communications, applications and platform design. We distinguish between mobility of the \emph{physical resource}, discussed here, and mobility of the logical applications, which we examine later.

Cloud data centers, obviously, are not mobile though their platforms can ease the mobility of data and applications among their geo-distributed data centers. However, spatial mobility at the edge layer is frequent through not universal. Edge mobility is exemplified in ubiquitous smart phones, connected vehicles, and drones, while they remain static in, say, traffic cameras and smart power meters~\cite{stojmenovic:fedcsis:2014,Loke2015TheIO}. Likewise, the Fog layer can also be manifest as a static or mobile resource platform~\cite{luan:arxiv:2016}. A Fog server can be installed at fixed sites such as a coffee shop or the airport, or on mobile vehicles such as taxi cabs or trains. Fig.~\ref{fig:mobility} captures these scenarios.

This mobility can have consequences on the link between the Edge and the Fog, or from the Edge to the public or private network, and these depend on the communication protocols used. However, we need to recognize that mobility can cause the network connectivity to be intermittent~\cite{shi2012serendipity}. This can cause a transient loss of access to data or compute resources between the Edge and the Fog or the Cloud, and the Fog and the Cloud. But, in technologies where the access is based on physical proximity, such as Bluetooth, Near Field Communication (NFC) or emerging line-of-sight technologies like Millimeter Wave, the disconnection can be permanent~\cite{vaquero2014finding}.

While we do not expect this ephemeral behavior for links between the Fog and the Cloud, such cases are possible in interaction among edge devices or with the Fog. For e.g., we use the notion of a ``Data Sherpa''~\cite{mishra:iot:2015} to transfer observed data from a static sensor (edge) to a mobile smart phone (edge) using Bluetooth when they are close by, and then use the smart phone to push the data to the Fog or Cloud. But the interaction between the two edges is opportunistic, not planned or repeatable~\cite{shi2012serendipity}. Similarly, users in a train can use their personal devices with the Fog servers in the coach to access media or make use of a high-speed Internet, but only while they are riding. Connected cars (Fog) can peer with nearby vehicles to share information on road condition and collaboratively navigate traffic~\cite{luan:arxiv:2016}. 
%
As a result, depending on the mobility of these two layers, the application and platform will need to be designed based on \emph{permanent, transient, periodic, or ephemeral connectivities} between the layers and within the layers which can determine the \emph{reliability of access} to data, storage, network and computing resources.

Lastly, applications may also relate spatial proximity with \emph{trust and context}. Fog layers physically close to the edge may be trusted by the edge applications to share sensitive data or for the Fog to offer services. This is already used to start cars using a ``key fob'' present in the car rather than a key in the ignition, or using NFC to share contacts or Bluetooth to pair devices. This can be leveraged more actively by applications for trusted interactions between the Edge and Fog layers. This closeness also offers context for the interaction and can be used to serve content relevant to the person from the Fog~\cite{yi:mobidata:2015}.






\section{Applications for Fog Computing}
\label{sec:apps}

In this section, we go in-depth into the characteristics and requirements of applications that are well-suited to different forms of Fog resources, in concert with the Edge and Cloud resources. We also offer detailed use cases on Fog applications.

\begin{figure}[t]
	\centering
	\includegraphics[width=0.6\columnwidth]{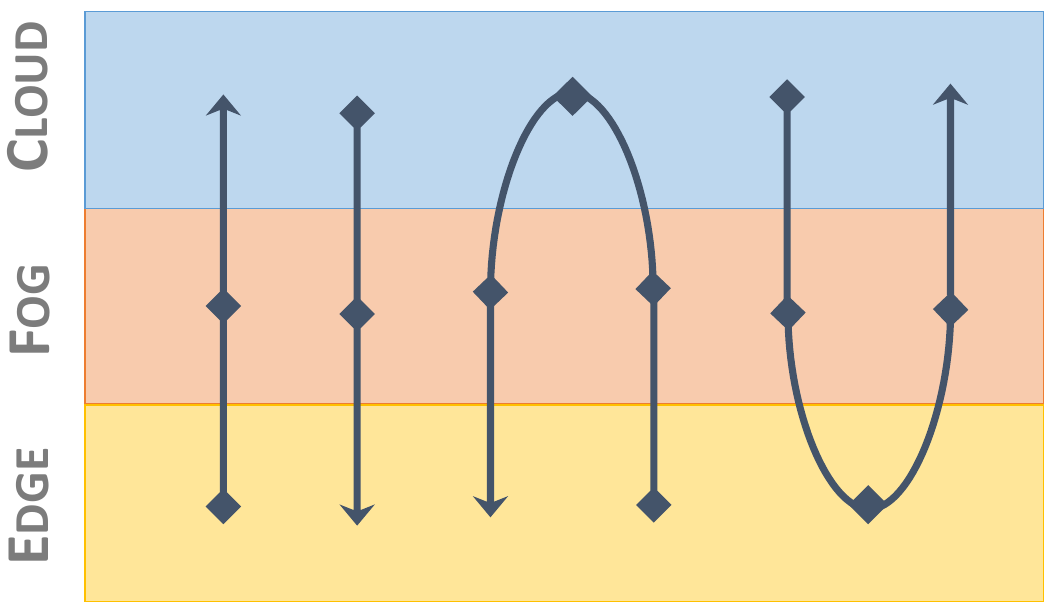}
	\caption{Logical mobility of applications between Edge, Fog and Cloud layers. Arrows indicate direction, diamonds indicate optional processing or staging at that layer. Process, Data and Events can move in both directions, with any of the layers as source or sink.}
	\label{fig:app:mobility}
	\vspace{-0.15in}
\end{figure}
\subsection{Mobility of Data and Analytics}
Similar to physical mobility of the resources, we can have logical movement of applications between Edge, Fog and Cloud layers. The entity that moves can be a \emph{process} (e.g., analytics, services), \emph{data} (e.g., video streams, device images) or \emph{event} (e.g., control signals, emergency notifications). Each have their own requirements from the resources.

Processes require appropriate execution environments in the resource layer and adequate compute and memory capacity. While virtualized Clouds offer the most flexibility in this regard with various VM images and sizes available, Fog is expected to have a similar flexibility using containers or hypervisors~\cite{yannuzzi:camad:2014,dasterdi:corr:2016}. Edge may be the most constrained in meeting dependencies, and the choices may be limited. Data movement requires bandwidth and storage to be available between the layers, and latency may be a concern as well. These sizes may range from Kilobytes for configuration files to Gigabytes to move high-fidelity observations to data archives~\cite{yi:mobidata:2015}. Events are a special category as they are by definition light-weight, and they serve as control signals that prompt an action. They typically require low latency transfer from source to sink, and the ability for the receiving layer to perform the action~\cite{stojmenovic:fedcsis:2014}.

There are four major \emph{logical mobility models} for the source, the sink, and their direction for this movement, as shown in Fig.~\ref{fig:app:mobility}. The first shows the logical entity move from Edge to Fog to Cloud, such as moving observation streams from Edge to Cloud for archival, with the Fog performing filtering or pre-processing~\cite{bonomi2012fog}. The second shows the entity move from Cloud to Edge, for e.g., flashing a new device image into a Pi deployed in the field, with the image cached in the Fog. Unlike these two ``open loop'' models, IoT applications often have a ``closed loop'' that consists of Observe, Decide and Act, and the last two modes indicate these feedback cycles. For e.g., the data source can be on the Edge, decisions are made at the Fog and/or Cloud, and the notification arrives back at the Edge for a response~\cite{yannuzzi:camad:2014,bonomi2012fog}. Similarly, the Cloud could provide a data stream to the Fog or Edge that respond back with a control message to enact a service call in the Cloud. Of course, subsets of these paths may be relevant for an application as well, e.g., start at the Fog, move to Cloud and then to Edge.

\subsection{Quality of Service (QoS) Characteristics}
There are several application QoS requirements that can drive the mix of resource layers to be considered. \emph{Latency and bandwidth} are two obvious factors for applications, and \cite{jin2012network} has earlier grouped applications based on their ability to withstand such delay. \emph{Inelastic applications} require real-time processing -- typically related to safety of humans or equipment such as autonomous vehicles, or near real-time processing for, say, applications consuming transient sensor streams~\cite{Bonomi2014}. The latency accumulates across each layer that the data passes through, and the Fog may offer a better pair-wise cumulative latency and bandwidth with the Edge and the Cloud, compared to directly between the two. \emph{Elastic applications} are delay-tolerant and designed for off-line or batch processing, such as surveys using drones for planning~\cite{Loke2015TheIO}.

The \emph{robustness} of the application becomes important as well. Mission-critical applications that operate City infrastructure such as power grids or traffic signaling may require deterministic network connectivity and computing capacity even during disasters, that may be better provided close to the Edge or Fog~\cite{stojmenovic:fedcsis:2014,yannuzzi:camad:2014}. On the other hand, if the robustness relies on instantaneous availability of large computing capacity or scalable services, say for supporting periodic events at stadiums, Clouds offer this flexibility provided redundancy of communication channels and data centers are built in.

Lastly, \emph{cost and user base} can play a major factor in the use of various resources. For applications that have a growing user community, scaling based on an elastic service-based model offered in the Cloud or the Fog allows them to pay proportional to their community size and revenue~\cite{bittencourt:pgcic:2015,satyanarayanan:2009}. Sometimes, applications coupled to consumer hardware, like autonomous cars and wearables, come with in-built (free) computing capability, on the device or a smart phone, that can (or should) be leveraged. Others like city utilities with a slow-changing population may wish to have captive resources self-managed at the Edge and Fog due to the scale that can be leveraged, and complemented by the Cloud~\cite{yannuzzi:camad:2014}. Sometimes, the resource availability itself may increase based on the demand from these applications, such as Cloud data centers or cell phone towers coming up in regions of revenue growth.

\subsection{Policies and Constraints}
\emph{Privacy and security} is a growing concern as the number of connected systems grow. There may be explicit corporate policies or national laws that do not allow applications to move data or services outside network or geographical boundaries~\cite{yannuzzi:camad:2014}. Here, the Edge and the Fog may be critical to meet performance needs as also legal compliance. On the other hand, Cloud data centers in general have stronger physical and digital security systems in place that may surpass what can be provided for widely distributed Edge and Fog devices that may operate in hostile environments~\cite{satya:pervasive:2013}.

IoT applications are often designed as \emph{Cyber Physical Social Systems (CPSS)}, that span data and software (cyber), infrastructure and devices (physical), and human interaction (social). In some cases, the application is intrinsically tied to specific devices (e.g., drones)~\cite{Loke2015TheIO}. In others, they may have explicit dependencies on computing infrastructure or resource capabilities. For e.g., a video classification may require model-training to take place on GPUs or FPGAs, which may be available only on the Edge or Fog. The application may also be coupled to a particular user who is mobile, and it will need to use logical mobility to move state and services to the resource that is physically closest to the user~\cite{bittencourt:pgcic:2015}. Consequently, some of the applications may be tightly-coupled or constrained to those cyber and physical systems.




\subsection{Application Use Cases}

\subsubsection{Urban Surveillance} 
There is a global push toward Smart Cities and India is poised to upgrade over 100 cities as part of a national initiative~\footnote{Smart Cities Mission, \url{http://smartcities.gov.in/}}~\cite{mishra:iot:2015}. As part of a \emph{Light-Pole Computing} pilot, several city blocks in Bangalore are installing video cameras, environmental sensors, Edge computing platforms (e.g., Raspberry Pi) and, at each city block, accelerated Fog computing platforms (e.g., NVIDIA Jetson TX1)~\footnote{Smart City test bed, Bharadwaj Amrutur, RBCCPS, 29 November, 2016, \url{http://www.rbccps.org/smart-city/}}.

Given the advances in deep learning, video surveillance in urban environments is essential not just for public safety but also as a proxy for ambient observations. For e.g., these analytics can be used to automatically identify parking violations, classify vehicles and people, generate visual summaries for consumption by safety agencies, and even blur regions for privacy~\cite{yi:mobidata:2015,jin2012network}. Training the neural network models for classification is computationally costly, and the source video streams at the edge have large sizes as well. Training may be required as frequently as every few hours to adapt to local conditions. Classification using trained models is faster, but even GPU accelerated Fog platforms can process only hundreds of frames per second.

Here, the training can be done in batch but moving the compute and data to the Fog saves bandwidth. The classification can move to the Edge or Fog, based on latency needs. At night, when women's safety is a concern, real-time needs trump daytimes use cases of detecting parking violations. As an example of a feedback control loop, the PTZ camera can be controlled by the analytics to automatically zoom into features of vehicles. Safety situations detected at the Edge can trigger an alert to spatially proximate smart phones, or notify officers through the Cloud~\cite{hong:mcc:2013}. The Edge and Fog platforms can also host other IoT applications from public utilities that tap into the video or sensor streams for event-analytics~\cite{ghosh:arxiv:2016}.












\subsubsection{Smart Power Grid} 
The \emph{Los Angeles Smart Grid} will serve over $4~Million$ customers in the largest public utility in the US~\cite{simmhan:cise:2013}. Net-connected smart meters observe power demand at households and industries and report them periodically back to the utility every few minutes. These run off 2G or P2P communication models. Further, SCADA systems at hundreds of city feeders collect high-frequency data on voltage and current quality across multiple phases, often at KHz rates.

\emph{Demand-response (DR) optimization} refers to shaping or shifting power demand to match supply capacity~\cite{aman:smartgridcomm:2015}. Here, liability and privacy concerns mean that demand-response and load control decisions happen at multiple-levels. The utility uses global but coarse-grained data to run demand forecasts and determine the curtailment level required~\cite{aman:tkde:2015}. It then notifies the customer gateway, say an Edge device for a consumer or a Fog device for a campus like USC, of the curtailment required and the discount (or penalty) pricing. These gateways then take local decisions to determine curtailment strategies and control, say a smart appliance or an electric car, or centrally change set-points of HVAC systems across campus buildings.

While the DR feedback loop can tolerate delays of several seconds or minutes, another smart grid application is one of \emph{state estimation} to determine the health of the distribution network. Here, computational models run over fine-grained power quality measurements from feeders to detect instability, and take rapid response on the Edge to avoid the whole network from being affected. There are distributed state estimation models that are well suited for Fog devices, with periodic updates pushed to the Cloud and global aggregates pushed back. As one can imagine, cyber-security concerns of such critical infrastructure eliminate the possibility of having multi-tenancy on these Fog resources.








\subsubsection{Drones for Asset Monitoring}








Lastly, drones offer a novel and emerging platform that highlight the role of mobility in Edge and Fog computing. Swarms of these autonomous platforms are being put to use for asset monitoring for gas pipelines or city infrastructure. Drones have on-board navigation and environmental sensors, visual/IR cameras, and sometimes even actuators (e.g., probes, grapples)~\cite{Loke2015TheIO}. They have wireless tethers with mobile or static base-stations, often on a line-of-sight, that act as the Fog layer. Upon return, these drones are charged by the base station and their data off-loaded before the next sortie. Static base stations will have high-speed connectivity to the Cloud, while mobile ones may rely on 4G.

The drones perform on-board real-time Edge computing for navigation~\cite{jin2012network}, combined with control signals from the Fog, which is required for its survival. Additional processing of observations for applications may depend on the spare computing capacity, battery levels and the elasticity needs of the application. The mobility of the drones and the base allows P2P mechanisms to be effected between the drones using high-speed millimeter wave line-of-sight communication or even between multiple bases. This can be used to move data or processes among them to conserve battery or reduce application latency, such as a mini-drones with constrained resources transmitting data to larger drones while in-flight~\cite{Loke2015TheIO}.
\section{Fog Computing Platform Abstraction}
\label{sec:platform}

Fog computing, even as a concept, is evolving and as a result, there is limited insight on what the \emph{application runtime and management platform} will turn out to be. They have the possibility of incorporating a service model similar to the Cloud as they are not resource-starved like the Edge, but may also inherit some requirements of mobility and performance-sensitivity seen in the Edge~\cite{bittencourt:pgcic:2015}. Here, we discuss how the system and application characteristics of Fog computing influence the features and abstractions offered by its platform. 


\subsection{Platform Management}
One of the key gaps in the realization of Fog computing is the lack of a platform ecosystem to design and run applications that can leverage the Fog, in conjunction with Edge and Cloud layers. There are two key parts to this: a \emph{fabric} for device management, and a \emph{platform} for application runtime management. 
Cloud computing fabrics expose every resource ``as a Service'', and this in part is a reason for its success~\cite{bittencourt:pgcic:2015,stojmenovic:fedcsis:2014}. Fabrics like OpenStack manage, schedule and instantiate VM images and instances, block and table storage, and networking capacity. Virtualization allows applications and Big Data platforms designed for desktop and cluster computing work equally well on Clouds.

Device management and application platforms on the Edge is a challenge. Edge computing is still done in either an \emph{ad hoc} manner, or tightly coupled to a smart phone platform like Android using sandboxed ``apps''. There are emerging IoT specifications like IETF CoRE, OASIS MQTT and oneM2M to enable management, data transfer and control of edge and embedded devices. IoT edge management software like Azure IoT Gateway, Amazon AWS Greengrass and VMware Liota are starting to be available. Application platforms like Apache Edgent and MiNiFi attempt to make logic deployment and migration on the Edge easier. 
However, this is still evolving.  

Fog computing is at an even earlier stage of platform maturity, but can gain from advances in both Cloud and Edge. Slimmed-down versions of Cloud fabrics could manage clusters of Fog devices, though they may have to operate in a wide-area network, while Fogs that are resource-light can make use of platform software for the Edge~\cite{Bonomi2014}. An IaaS model for Fog resources would offer the most flexibility, but interoperability of APIs becomes even more important compared to Clouds -- large-scale deployment of Fog resources will take time, and utilizing resources from different providers, like ``roaming'' using cell-phone providers, is essential~\cite{luan:arxiv:2016}. Containerization like \texttt{lxc} and Docker are likely to be more favored for Fog rather than hypervisors, though advances like Ubuntu's \texttt{lxd} may make Linux VMs responsive while also offering capabilities like live-migration which can be invaluable for mobile Fog resources~\cite{yannuzzi:camad:2014,bittencourt:pgcic:2015}.






\subsection{Application Composition}

The execution environment offered by Edge, Fog and Cloud is inherently distributed, and the application space is vast as well. There are many application definition models used in such scenarios such as control flows, data flows, event-driven models~\cite{seda}, and so on. Directed Acyclic Graphs (DAGs) are popular for capturing flow dependencies in complex distributed applications, and are widely used for Cloud and Edge computing~\cite{shi2012serendipity}. At the same time, latency sensitive applications may prefer an event-driven model that react rapidly to changing situations~\cite{yi:mobidata:2015}. 

The ability to encode priorities among tasks and temporal processing guarantees for events will be crucial as well for mission-critical systems. Besides latency, energy and costs are likely to be important parts of the application specification~\cite{dasterdi:corr:2016,Bonomi2014}. Geo-fencing policies may also limit the movement of data or services to specific resources.

The data flowing between loosely coupled parts of an application may be based on \emph{streams, micro-batches or files}, again depending on the latency and processing costs. While streams offer a low-latency data transfer for sensor observations, micro-batches have grown popular off-late as they offer a balance between managed latency and higher throughput by amortizing per-tuple overheads~\footnote{Apache Flink, Apache Spark Streaming}. Files are well suited for defining batch processing applications. The ability to define specialized data structures, compression and transport mechanisms for distinct stream types such as audio and video may be necessary as well.

\emph{State} is likely to play a key role in such applications, and the composition models need to offer a way to include it as a first-class entity. The state may be associated with a user, device or session, and it will need to migrate across space (Edge/Fog devices) and time, and offer a context for execution~\cite{bittencourt:pgcic:2015}. Another unique element of such application is the role of \emph{spatial proximity} or location-awareness in determining actions. For e.g., proximity of two devices bay trigger an action, such as a data transfer, and this may be dictated by their geo-location as well~\cite{Loke2015TheIO}. Some may wish to explicitly couple the application definition with the placement of their constituent tasks, either for performance reasons (e.g., sensing task on edge, processing task on Fog) or for spatial locality to Edge~\cite{hong:mcc:2013}.

\subsection{Orchestration and Coordination}


The application definition needs to be scheduled and coordinated in order to meet various QoS goals such as latency, energy and monetary constraints. This coordination can be done using different strategies. Three common orchestration model that are relevant in such a multi-layered and distributed resource environment are \emph{centralized, hierarchical} and \emph{peer-to-peer (P2P)}. We also distinguish between \emph{scheduling decisions}, the \emph{flow of control signals} and the \emph{flow of data}, and different coordination models could be applied to these.

\emph{Centralized} orchestration has a single service, either per application or for the platform, that is located in one of the three resource layers, and is responsible for making scheduling decisions, and coordinating the transfer of control signals and/or data items. For e.g., in \cite{dasterdi:corr:2016}, a central resource management layer determines the best resources to schedule the incoming tasks using the monitoring information from the Edge, Fog and Cloud resources. This is simple to design but can suffer from high latencies and transfer costs, and is a single point of failure. While this orchestrator often runs in the Cloud (to coordinate across edge devices) or the Edge (to interact with different Cloud services), the Fog layer could offer a sweet-spot for such a centralized coordinator. For e.g., when a Fog sits at the gateway between private and public networks, it is in the sole position of being able to judge the behavior of resources in the Edge (private) and the Cloud (public)~\cite{aazam:2014}. 

A \emph{hierarchical} architecture is a generalization of the centralized model, and allows only vertical communication of data and controls to take place between adjacent layers. This is a natural fit for Fog computing as it leverages both the bandwidth and latency benefits of the Fog layer in accelerating these flows, as well as the compute benefits closer to the observation source~\cite{yannuzzi:camad:2014,stojmenovic:fedcsis:2014,yi:mobidata:2015,stojmenovic:fedcsis:2014,dasterdi:corr:2016,hong:mcc:2013}. Often, the Cloud forms the root of this tree and is used for global data aggregation and coordination. Local data analytics is delegated to Cloudlets and further to the edge devices. This allows a federated behavior that has shown to scale. 

\emph{P2P} is a form of distributed coordination that avoids a single point of failure. Here, peers in the same Edge or Fog layer  
can pass control and data directly among each other~\cite{vaquero2014finding}. The horizontal communication channels may initially be setup by an entity that has a global picture of the resources. This is typically done at the Cloud or the Fog, or one of the edge devices that serves as a leader. There simple component based models for composing and executing P2P applications, as well as complex ones that use Distributed Hash Table (DHT) to maintain an overlay network over peers that frequently enter and leave the system. 
For example, in~\cite{luan:arxiv:2016} the fog devices peer amongst each other for content delivery and service provisioning for improving the performance.
Similarly ~\cite{stojmenovic:2014} consider a hierarchical architecture where proximate smart traffic lights coordinate among themselves to send green traffic wave or alert the approaching vehicles.


In a \emph{hybrid} model, there are no strict limitations on the flow of control or data flows, and all layers are seen as having resources of heterogeneous characteristics. While there can be interconnections among resources within each layer (Cloud, Fog, Edge), communication can also take place vertically~\cite{bittencourt:pgcic:2015}. This can help in different situations. 
If a higher layer is not reachable due to network failure or resource unavailability, then the tasks can be accomplished within the lower layers itself though with a graceful degradation of the QoS~\cite{madsen2013reliability}.
Similarly, if a Fog layer is unable to reach an edge device, it can replicate the last state of the application running on it to a different edge device. Such hybrid strategies are likely to require more complex coordination, but can potentially improve the resilience of the application.

\subsection{Managing Dynamism and Mobility}

A Fog computing platform has to be responsive to various forms of \emph{dynamism}. As identified before, there can be resource mobility at the Edge and Fog layers. The applications may also impose requirements on logical mobility of processes and data. Further, there may be changes in the data generation rates, network behavior or energy levels of batteries that requires reactive strategies~\cite{shi2012serendipity}. Thus, the runtime environment should offer monitoring capabilities to determine when such adaptation is required, and provide transparent mechanisms for enacting these changes~\cite{dasterdi:corr:2016}. This may even require \emph{changing the coordination strategy} from, say, centralized to P2P.

Such \emph{migration} may involve moving just state from one service to another, moving an entire application and its dependencies, or moving a VM or container as a whole.  For e.g., \cite{bittencourt:pgcic:2015} migrate the user's data from one Cloudlet to another as per the mobility of the user in order to minimize latency. 
Depending on whether the Edge and or the Fog resources are physically mobile, we may need to preemptively off-load to a different resource or layer when proximity or connectivity is going to be lost, or periodically save state to the Cloud~\cite{shi2012serendipity}.


For real-time applications or those that have a closed control loop are particularly sensitive to mobility and require careful platform design. Often, these preclude the use of mobile resources, or retain the decision logic in a single resource. For e.g.,~\cite{sadeghi:hipc:2016}, perform video analytics to blur the scenes of movies based on the mental state of the user with a medical condition, but offload the signal processing task to a static Fog server. 
%
Batch applications offer the most flexibility and can efficiently make use of mobile Edge and Fog resources 
to opportunistically offload the tasks or data 
as and when resources are available. For e.g., mini drones may be able to offload their data to larger drones or to a Fog server at the base-station when they come within range~\cite{Loke2015TheIO}.

\section{Related Work}
\label{sec:related}
There have been several early papers that conceptualize the idea of Fog computing and Cloudlets. Cisco~\cite{Bonomi2014} popularized the notion of Fog computing as a complementary computing layer from Cloud that is driven by distinct applications that require low and predictable latencies. 
They discuss the importance of multi-tenancy on the fog layer while ensuring mission-critical operation, using a ``Foglet'' as a software agent on the Fog layer to manage the local operations and interface with the Cloud.
We go beyond this, and discriminate between edge devices and the Fog layer, and discuss the interaction between all three layers. 
We also include mobility as first class entity, not just at the Edge but also at the Fog layer. 
Further, while a centralized, distributed or hierarchical models are mentioned, we contrast between control flows and data flows that may each use a different interaction model. 

\cite{satya:pervasive:2009} introduced the concept of Cloudlets as resource-rich infrastructure that is within one network hop of the mobile edge device, consistent with the concept of Fog Computing, as a second-level data center that is proximate to mobile devices at the edge~\cite{satya:comm:2015}. 
They offer examples of the latency benefits for augmented reality and face recognition applications, 
nut these treat Cloudlets more as a ``CDN in reverse'' to avoid latency and bandwidth costs. 
They use VMs used to deploy and manage applications, either through active migration or of state overlays over existing VM images. They also discuss the business model for Cloudlets, spanning between retain owners and service providers.

\cite{vaquero2014finding} attempt to define Fog computing, but views it from the narrow prism of network management and connectivity. 
They also tabulate various features and challenges in brief. Our interest in this paper is from the application, platform and middleware perspective. We further consider the system features and the applications that benefit or need Fog.


Some early research investigates platform and application models for Fog computing.
\cite{hong:mcc:2013}~propose a programming model for composing applications that run across mobile (Edge), Fog and Cloud layers as a Platform as a Service (PaaS). They offer a multi-way 3-level tree model where the computation is rooted in the Cloud, resources are elastically acquired in the Cloud and Fog layers, and communication is possible between Cloud and Fog, or Fog and Edge. 
A strictly hierarchical model while simple, limits the flexibility in application composition. 
Their example applications do not consider a role for the Cloud either, though their APIs support it. This degenerates to a client-server model between the edges and their Fog parent. Further, the interactions between edge devices and Fog layer should be actively used as well, rather than only vertical interactions. Lastly, while mobility of the edge is discussed, this does not consider when the Fog can be mobile as well. 

The role of virtualization in enabling Cloud computing is discussed in~\cite{bittencourt:pgcic:2015}, and they see a similar role for Fog computing as well. They conceive of a VM encapsulating all necessary dependencies for an edge application or user to be hosted on a Cloudlet within 1 hop of the edge, with this VM moving with the edge user to remain at 1-hop distance. This virtualization architecture should expose API for the developers to offload data and processing, synchronization of data among replicas, discovery of Cloudlet resources, and the migration of the VMs across Cloudlets. 
However, we take a broader platform view and discuss the possible architectural designs for the Fog.

Other literature examine specific applications that benefit from the Fog layer. \cite{dastjerdi:computer:2016} discuss the role of Fog computing for IoT as Edge computing alone is inadequate to deal with multiple IoT applications. Fog helps with coordination of distributed edge devices and uses Cloud resources. They consider Sense-and-actuate and stream processing as two programming models. But we argue there can be more diverse application composition and coordination models. 

\cite{yannuzzi:camad:2014} discuss the need of Fog computing for real-time applications such as sensing of gas pipelines, smart agriculture and control systems inside factories. They consider a hierarchical architecture where the data is analyzed and processed at one level and then sent to the higher level for further aggregation and analysis. Other possible architectural designs and types of applications are missing. 
Similarly,~\cite{stojmenovic:fedcsis:2014} also discuss the need of Fog computing for smart city applications. 
Further they have looked into some of the privacy and security issues that are possible in fog computing. What is lacking in these is a broader examination of application characteristics rather than examples that motivate the Fog, which we address.

\cite{luan:arxiv:2016} attempt a similar effort such as our paper, but limit their exercise to mobile devices rather than edge devices at large. 
They recognize that the Fog can be static or mobile, similar to entertainment systems in vehicles. 
They highlight that Fog can deliver location-aware content unlike Cloud, but this reduces the value of a Fog to that of a CDN, only closer to the edge. However, it is important to note that many Cloud services are capable of using geolocation using network or device GPS to offer location sensitive information. Rather, we state that the physical proximity also offers a certain degree of trust by the client of the Fog, and the ability to host rich interactive services, not just content. 
Their discussion on research problems delves more on the networking and communication between Mobile/Cloud and Fog, and between Fogs rather than on application and middleware.

\cite{yi:mobidata:2015} offer a brief survey of Fog computing concepts, in which they include both resource poor devices (which we refer to as edge) and resource rich Cloudlets and Cisco's IOx. They highlight Augmented reality, Content delivery and Mobile Data Analytics as three motivating applications to reduce the latency delays, and reduce bandwidth costs. They do not offer any analysis of Fog computing architecture or platform dimensions. As before, network management using Network Virtualization and SDN appear as technical issues to tackle. They identify the need for QoS, programming APIs, resource provisioning, security and billing as aspects to address.

\cite{stojmenovic:2014} present a survey and discuss a hierarchical Fog-based architecture. We discuss several alternative architectural and coordination designs rather than a one-size fits all. They too consider IoT applications 
but fail to present a taxonomy of the application characteristics that can benefit from Fog as we do.


\section{Discussion and Conclusions}
\label{sec:discuss}
\subsection{Challenges}
There are several interesting problems that arise in the context of Fog computing, and addressing these can pave way for the technical feasibility.
\begin{itemize}
	\item \emph{Programmability:} It should be easy for the application developers to run their applications on the Fog. This will require the interface and the programming model to hide the complexities of Fog computing. Fog components should also allow for application-specific customization and optimizations~\cite{dastjerdi:computer:2016,vaquero2014finding}.
	
	\item \emph{Predicting demand:} One of the major challenges for deployment of Fog devices is to predict the users' demand and accordingly determine the quantity of resources required at different locations~\cite{luan:arxiv:2016}. 
	Thus it is a challenge to optimally place the fog devices at appropriate locations so as to serve the requests of nearby mobile users.
	
	\item \emph{Power and Network Consumption:} Tasks in Fog devices are distributed across multiple locations compared to more centralized Clouds~\cite{dastjerdi:computer:2016}. Due to the distributed nature of computations, captive local power supply and network connectivity will be important. This can be a challenge for developing countries without $24\times7$ power supply. Task placement may need to be optimized to minimize the power consumption, or leverage renewables like Solar.
	
	\item \emph{Where to push the tasks?} One has to decide how to partition an application and distribute the tasks among the three layers. This depends on the application requirements~\cite{yi:mobidata:2015}. 
	
	\item \emph{Security and fault tolerance:} Since Fog resources are geographically distributed and can have multiple service providers, ensuring application and data security is a key challenge~\cite{stojmenovic:fedcsis:2014,dasterdi:corr:2016}. Also, one needs to plan how to handle failures of Fog devices. When the nearby Fog devices are inaccessible, users may need to push their applications to the remote Fog or even the Cloud, which may impact the performance of the applications.
	
	\item \emph{Fog providers and billing:} Fog computing services may be provided by Internet, Cloud and Cell-phone service providers, retail merchants, or emerging ``gig economies'' like taxi or hotel aggregators. Pricing and billing remains a challenge in terms of sustaining a commercial ecosystem or value added services~\cite{yi:mobidata:2015}. 
	Moreover there are possible billing models such as \emph{consumption based} where the users are billed as per their usage or \emph{subscription based} where the users pay a fixed price monthly and can use the Fog network-wide~\cite{bittencourt:pgcic:2015}. 
\end{itemize}

\subsection{Reality Check}
It is still early days for Fog computing~\cite{vaquero2014finding}. Growth of fog has no where near kept up with the growth in edge devices. So the pain-points or gaps of a lack of fog computing have not reached a threshold. These are starting to emerge more in vertically integrated ``private'' scenarios rather than horizontal, reusable ``public'' ones, much in the way of Cloud data centers being private to Amazon, Google and Microsoft, before the business model fell into place for commercialization~\cite{yi:mobidata:2015}.

We do not have large scale deployments of Fog Computing or its commercial operation in a pay-as-you-go on-demand model like public Clouds currently available. But the need for it is growing and we can foresee several existing infrastructure operators evolving to offer Fog services as well. Two prime contenders are \emph{operators of cell-phone towers} who tend to have captive power, communications and space, and can complement this with computing resources~\cite{satya:comm:2015}, and \emph{taxi aggregators} like Uber and Ola who are already offering value added services such as WiFi and grocery delivery, and can incrementally extend this to a Fog server hosted in the boot of their cab. At the same time, emerging infrastructure such as \emph{Smart Cities} will see the organic growth of the Fog layer, initially integrated as part of a vertical domain such as smart power grid or smart transportation (e.g., collocated with a transformer or traffic light, at each city block), and later offered as a horizontal resource shared by multiple city or commercial services~\cite{stojmenovic:fedcsis:2014,bonomi2012fog,Bonomi2014}. Such ``light-pole'' sensing and computing is already starting, as mentioned earlier.

At the same time, not all mobile or IoT applications will necessarily need or even benefit from having a Fog Computing layer. For e.g., as Smart Power Meters are rolled out across Los Angeles, the largest public utility in the US, these meters record the electricity usage data at $15~min$ granularity to support demand-response optimization decisions that are taken periodically~\cite{simmhan:cise:2013}. Even with $\approx4M$ customers, this works out to about $400~GB$ of data collected each day (assuming a $1kb$ payload per observation), or about $4~GB$ of distributed data collected per $15min$ interval. This is meager, even for a mega-city like Los Angeles, compared to the $24~GB/sec$ streamed by Netflix during peak hours~\footnote{Evolution of the Netflix Data Pipeline, February 15, 2016, \url{http://techblog.netflix.com/2016/02/evolution-of-netflix-data-pipeline.html}}. While there are other Smart Grid applications that may require higher sampling rates and data sizes (e.g., realtime analytics over Phasor Measurement Units (PMU) data~\cite{interrante:hpcc:2012}), these are still evolving. So not every Smart City or IoT application requires low latency or high bandwidth or complex analytics.

The concept of Utility computing itself existed long before Cloud computing in its current form emerged as a feasible business model, with Grid computing having meanwhile been tried, less successfully, in the academic community~\cite{youseff:2008,madsen2013reliability}. Likewise, sensor networks and pervasive computing concepts existed a decade before the technologies converged to offer IoT at massive scales. Likewise, the reality is that the gestation period for Fog computing may be anywhere from a couple of years to a decade before it matures into a sustainable technology and business model. In fact, one of the early analogies to Fog computing was the \emph{Google Search Appliance}, which offered on-premises search capability for enterprises, with the Cloud only for support. But this died a natural death with network bandwidth improving and the Cloud capabilities like deep-learning far out-stripping what was possible at the GSA~\footnote{\url{http://www.theregister.co.uk/2016/02/10/google_quits_search_appliances/}}. So the eventual outcome of Fog computing may indeed be different from what one might envision, but offering dimensions for characterization allows this eventual manifestation to fall at some point in the spectrum.

\bibliographystyle{plain}
\footnotesize{
\bibliography{paper}
}

\end{document}